\documentclass[12pt,a4paper,pdftex]{article}
\usepackage{graphicx}
\usepackage{times}
\usepackage{upgreek}
\usepackage{amsmath,amssymb}
\usepackage{url}
\title{\bf Asteroseismology as a new window \\ 
to statistics of binaries
}

\author{Hiromoto Shibahashi$^1$\thanks{email: shibahashi@astron.s.u-tokyo.ac.jp},  
\hspace{0.1cm} Simon J. Murphy$^{2}$
\vspace{0.5cm}\\
\normalsize $^1$ 
Department of Astronomy, School of Science, University of Tokyo, Tokyo 113-0033, Japan\\ 
\normalsize $^2$ Sydney Institute for Astronomy, School of Physics, University of Sydney, NSW 2006, Australia 
}
\date{\mbox{}}
\begin{document}
\maketitle
\setcounter{page}{1001}
\pagestyle{plain}
    \makeatletter
    \renewcommand*{\pagenumbering}[1]{%
       \gdef\thepage{\csname @#1\endcsname\c@page}%
    }
    \makeatother
\pagenumbering{arabic}

%
%
\def\bull{\vrule height .9ex width .8ex depth -.1ex}
\makeatletter
\def\ps@plain{\let\@mkboth\gobbletwo
\def\@oddhead{}\def\@oddfoot{\hfil\scriptsize\bull\quad
"How Much do we Trust Stellar Models?", held in Li\`ege (Belgium), 10-12 September 2018 \quad\bull}%
\def\@evenhead{}\let\@evenfoot\@oddfoot}
\makeatother
%
%
\def\beginrefer{\section*{References}%
\begin{quotation}\mbox{}\par}
\def\refer#1\par{{\setlength{\parindent}{-\leftmargin}\indent#1\par}}
\def\endrefer{\end{quotation}}
%
%
{\noindent\small{\bf Abstract:} 
More than 340 non-eclipsing binaries of A-F stars as primaries at intermediate periods (100\,-\,1000\,d) were newly found by uninterrupted photometry with ultra high-precision taken over 4\,yr by {\it Kepler} space mission via the phase modulation of pulsating stars, and their orbital parameters, which were difficult to measure with conventional methods and were very incomplete, were then determined.   
This asteroseismic finding of binaries tripled the number of intermediate-mass binaries with full orbital solutions, and opened a new window to statistics of binaries.
We outline the methods of finding non-eclipsing binaries by photometry and of deriving the orbital parameters, and demonstrate their validation. 
Statistical study of the newly discovered binaries along with the known spectroscopic binaries near A-F main-sequence are then presented, converting distribution in the mass-function into the mass-ratio distribution with the help of Abel-type integral equation.  
}
\vspace{0.5cm}
\section{Introduction}

The study of binary stars is a principal source of stellar fundamental parameters, which form the basis of our understanding of stellar structure and evolution. The primary observational data for binary stars are astrometry in the case of visual binaries, the photometric light curves in the case of eclipsing binaries, and the radial velocity curves in the case of spectroscopic binaries. 
These data have to be collected as time series covering the orbital phase of each binary system, often obtained one by one from ground-based observing sites with much time and effort from observers.
A large amount of observing time is required, often on large telescopes for fainter stars. This is the bottle-neck to binary studies, and as a consequence, the high-precision masses and radii for stars were known only for a few hundreds of nearby stars, even though binary systems are ubiquitous (Moe \& Di Stefano 2017).

The importance of studying binary stars is not limited to precise measurement of stellar masses and radii. The statistical study of the mass-ratio distribution and orbital parameters are also important for understanding the physics of binary systems, including binary star formation processes, dynamical interactions between binary components, the evolution of binary stars including mass transfer, and so on. 
However, such statistical studies are few in number. Examples include 
Duquennoy \& Mayor (1991)  
and 
Raghavan et al. (2010) 
for solar-type primaries, and 
Sana et al. (2012)  
and 
Kobulnicky et al. (2014) 
for binaries with O/B primaries, but even in the most comprehensive reviews 
(Duch\^{e}ne \& Kraus 2013, Moe \& Di Stefano 2017)
intermediate-mass (A/F) stars are mostly left out because of a lack of data. It has been practically difficult to homogeneously sample a large number of binary systems, since each aforementioned methodology for collecting observations is favourably applied to its own preferable cases. 
Visual binaries are necessarily widely separated and consequently have long periods. Also, since faint companions are hard to see, systems with small mass ratios are hard to find. 
On the other hand, eclipses are geometrically unlikely for long-period binary systems, hence most eclipsing binaries discovered so far have short periods.
Similarly, since the Doppler shift of spectral lines due to orbital motion is small in the case of long periods, spectroscopic binaries are found preferably in short-period orbits.

Starting with the Canadian {\it MOST} mission (Walker et al. 2003), through ESA's {\it CoRoT} (Auvergne et al. 2009), NASA's {\it Kepler} (Koch et al. 2010) and {\it TESS} (Ricker et al. 2015) missions and the international {\it BRITE}-Constellation (Weiss et al. 2014), space-based photometry with extremely high precision over long time spans has led to a drastic change of this situation. This pedigree will be further taken over by the ESA mission {\it PLATO}\footnote{http://sci.esa.int/plato/59252-plato-definition-study-report-red-book/}.
The {\it Kepler} mission has been particularly transformative. {\it Kepler} monitored over 190\,000 stars almost continuously over its original four-year lifetime, and detected some variability in almost all stars. Tens of thousands of pulsating stars were newly discovered, along with hundreds of eclipsing binaries 
(Kirk et al. 2016). 

Using the {\it Kepler} photometry, 
Shibahashi \& Kurtz (2012) 
and later Murphy et al. (2014)  
devised techniques to measure the orbital motion of binary stars using their oscillations. 
Similar methods of using photometry to find binaries have been developed by Koen (2014) and Balona (2014).
Application of these techniques to thousands of stellar light curves of unprecedented quality led to the discovery of hundreds of new binary systems 
(Murphy et al. 2018),  
particularly at intermediate periods and for intermediate-mass stars where binary statistics were least complete. Importantly, the majority of these new non-eclipsing binaries would not have been detectable by other techniques, because the primaries are rapid rotators, making spectroscopic radial velocities (RVs) difficult to obtain, and their intermediate periods exclude their detection by other methods.

In Section\,2, we outline the method of finding non-eclipsing binaries by photometry.
A summary of the processes of deriving orbital parameters is given in Section\,3 and
validation of the method is given in Section\,4.
The method was applied to $\delta$\,Sct pulsators in the {\it Kepler} field by Murphy et al. (2018), and the results are summarised in Section\,5.  
A mathematical procedure to derive the mass-ratio distribution is then described in Section\,6, and the results are shown in Section\,7.
Finally, in Section\,8, we briefly describe a possibility of unveiling stellar-mass, X-ray-quiet black holes  lurking in binaries. 

A wide-ranging review on pulsating stars in binary systems was recently presented by Murphy (2018) with a comprehensive list of references. We refer readers there for further details.

\section{Method of finding non-eclipsing binaries by photometry}

Binary orbital motion causes a periodic variation in the path length of light travelling to us from a star. Hence, if the star is pulsating, the pulsation phase periodically varies with the orbital motion. 
In the past, the light-time effect on the observed times of maxima in luminosity, which vary over the orbit, was utilized to find unseen binary companions (the so-called $O-C$ method; see, e.g. Sterken 2005, and other papers in those proceedings). 
An example with great success is the discovery of the first binary pulsar by Hulse \& Taylor (1975), who were honoured with the 1993 Nobel prize for their discovery that opened up new possibilities for the study of gravitation in the relativistic regime.
Around the same time, the high-amplitude $\delta$ Sct star SZ\,Lyn was discovered to be in a 3.14-yr binary based on deviation of the photoelectrically observed ephemeris from the calculated values (Barnes \& Moffett 1975). The discovery was later verified with radial velocities (Moffett et al. 1988).
Such a pulse timing method works well in the case of stars pulsating with a single mode, since the intensity maxima are easy to track and any deviations from precise periodicity are fairly easy to detect. However, when the pulsating star is multiperiodic (e.g., Silvotti et al. 2007), as in the case of most {\it Kepler} objects, or when the star is in a multiple system (e.g., Wolszczan \& Frail 1992), the situation is much more complex.

Let us consider a star pulsating with multiple modes labelled by $\{k\}$ in a binary. The luminosity variation at time $t$ is then given by
\begin{equation}
	\Delta L(t) 
	= 
	\sum_{k} A_k\cos 
	\left[ 2\uppi\nu_k \left\{ t-{\frac{1}{c}}\int_0^t v_{\rm rad}(t')\,{\rm d}{t'} \right\} +\phi_k \right],
\label{eq:01}
\end{equation}
where $A_k$, $\nu_k$ and $\phi_k$ are the amplitude, the frequency and the phase at $t=0$ of the mode $k$, respectively,  $c$ is the speed of the light and $v_{\rm rad}(t)$ denotes the radial velocity due to the orbital motion of the pulsating star,\footnote{The so-called $\gamma$ velocity of the binary system (its common motion through space) cannot be measured here, unless the rest-frame oscillation frequency is known $a~priori$.} where the epoch is the time at which the star passes the nodal point directed towards us.
By convention, the radial velocity is taken as positive when the star is receding from us.
The second term in the curly bracket on the right-hand side is the time delay caused by the light travel time effect. 
For each frequency $\nu_k$, this time delay manifests itself as a periodically varying phase shift in the form of the product with the intrinsic angular frequency $2\uppi\nu_k$.
The light-arrival time delay is hence measurable by dividing the observed phase variation  by the frequency.
If multiple oscillation frequencies show the same periodic time delays, the origin can be attributed to the binary orbital motion.

The instantaneous frequency is regarded as the time derivative of the phase, and the Doppler shift of the frequency due to the orbital motion is given by $v_{\rm rad}(t)/c$, which is typically of the order of $10^{-3}$ or less.
The wider the orbit, the smaller the Doppler shift of the frequency.
Conversely, the amplitude of phase variation caused by the orbital motion is given by, as seen in equation (\ref{eq:01}),
the product of the intrinsic angular frequency $2\uppi\nu_k$ and the time integral of $v_{\rm rad}(t)/c$, 
which means the light travel-time across the orbit divided by the pulsation period, hence
wider orbits and higher pulsation frequencies give larger light travel time effects.

{\it Kepler}'s four-year monitoring allowed binary stars with orbital periods as long as $\sim 4$\,yr to be studied using this effect.  
That is, the searched orbital period range was significantly longer than that of typical spectroscopic surveys. In addition, nearly 200\,000 stars were simultaneously monitored, leading to hundreds of new discoveries and opening a window to a statistical study of binary stars and their orbits.

It should be remarked here, however, that this method is restricted to stars with coherent pulsations.  
Solar-like oscillations are widely seen in late-type stars and red giants. However, they are stochastically excited by turbulent motion in the convective envelope. Consequently, phase coherence is not sustained, ---that is, $\phi_k$ is not kept constant, hence the pulsation phase varies stochastically and the present method does not work well for solar-like oscillators.

\section{Deriving orbital parameters of binaries by photometry}

\begin{figure}[t]
\centering
\includegraphics[width=0.7\linewidth]{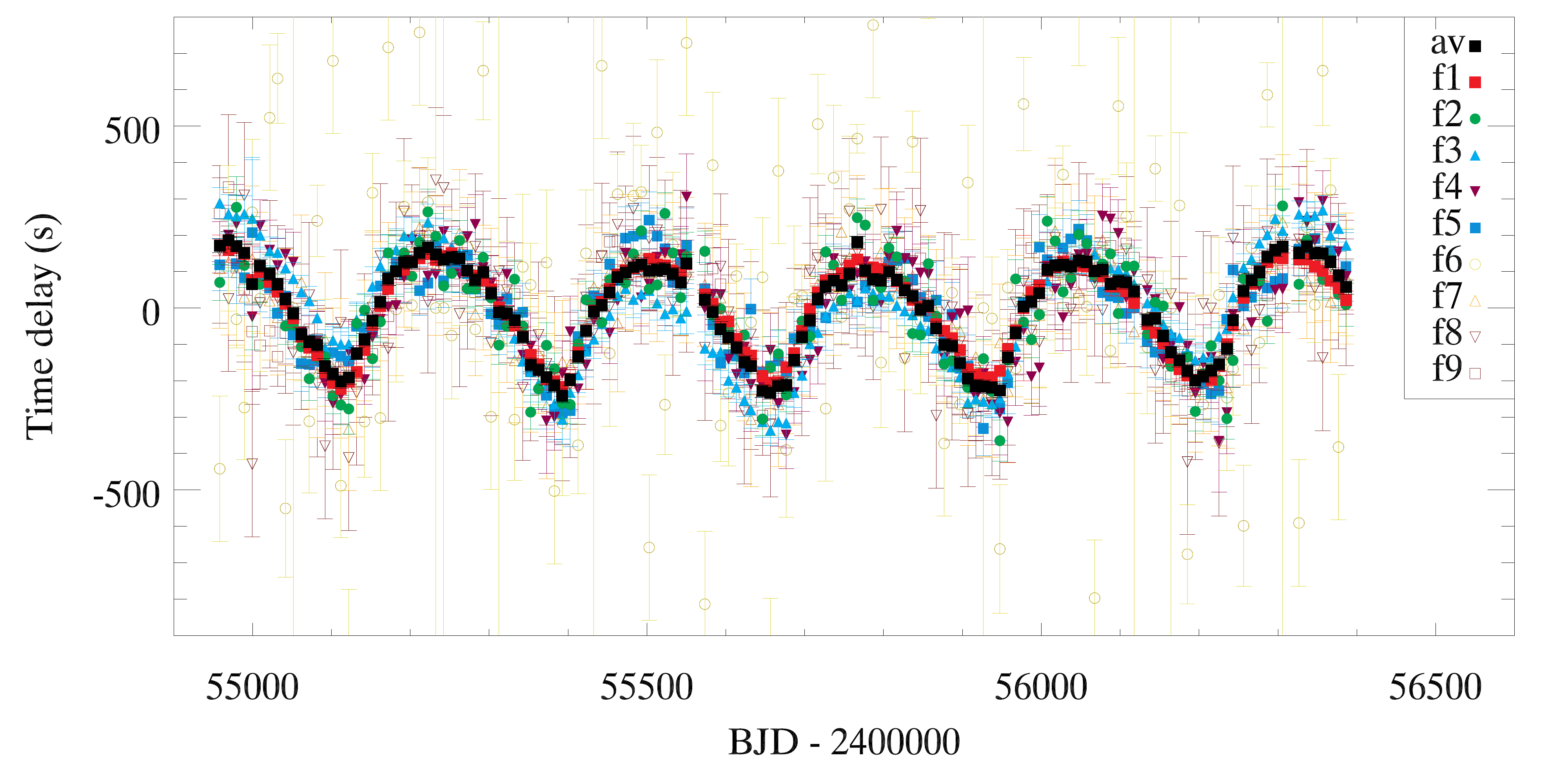}
\caption{An example of time delay curve (KIC\,9651065) using nine different pulsation modes. The weighted average is shown as filled black squares. Adopted from Murphy \& Shibahashi (2015).
\label{fig:01}}
\end{figure}

For each pulsating star, we first take a Fourier transform of the light curve taken over the full observational time span to determine the pulsation mode frequencies $\nu_k$ having the highest peaks in the p-mode range.
High-order p modes in the super-Nyquist frequency range should be correctly distinguished from aliases (Murphy, Shibahashi \& Kurtz 2013, Shibahashi \& Murphy 2018).
The light curve is then divided into shorter segments and the phase of the selected modes is measured in each segment. The segment size should be chosen so that the selected modes can be clearly distinguished and the pulsation phase can be well determined in each individual segment.
Once pulsation phase variation is obtained in this way as a function of time, it is converted to time delays by dividing the angular frequency of each peak.
It should be verified that the time delays deduced from different pulsation modes are in agreement with each other within error bars.
Weighting by the phase uncertainties, we finally deduce a time series of weighted mean time delays.
Figure\,\ref{fig:01} shows an example of time delay curve obtained from the {\it Kepler} photometry data, where the time delays vary periodically with the binary orbital period. 
The time delay curve thus deduced immediately provides us with qualitative information about the orbit.
Deviation from a sinusoid indicates that the orbit is eccentric.  The pulsating star is farthest from us when the time delay reaches its maximum, while the star is nearest to us at the minima. The difference between the maximum and the minimum time delay, measured in seconds, gives the projected size of the orbit in units of light-seconds. The sharp minima and blunt maxima in Fig.\,\ref{fig:01} indicate that the periastron is at the near side of the orbit. Fast rise and slow fall reveals that periastron occurs after the pulsating star reaches its nearest point to us.
 
Further careful analyses provide us with quantitative information about the orbit.
With the help of celestial mechanics, the time delay, denoted hereafter as $\tau(t)$, is analytically written with the orbital elements as 
\begin{equation}
	\tau(t) 
	=
	-{{a_1\sin i}\over{c}} {{1-e^2}\over{1+e\cos f}} \sin (f+\varpi),
\end{equation}
where, $a_1\sin i$ denotes the projected semi-major axis of the orbit of the pulsating star, $e$ is the eccentricity, $\varpi$ is the angle between the node and the periastron, and $f$ is the true anomaly, which specifies the temporal position of the star on the orbit for a given set of the orbital frequency $\nu_{\rm orb}$, the eccentricity and the time of periastron passage $t_{\rm p}$.
The next task is then to deduce the set of orbital parameters, 
$(a_1\sin i, e, \varpi, \nu_{\rm orb}, t_{\rm p})$, 
from the observed time delay curve.
It should be noted here that our practical process of dealing with the time delay averaged over the divided short segment causes the maxima and minima of the deduced time-delay curve to be less sharp than the real, continuous one that would be obtained instantaneously. Hence, this smearing effect due to undersampling should be corrected properly when the orbital elements are deduced. This process of obtaining the orbital elements has been described by
Murphy \& Shibahashi (2015) and by Murphy, Shibahashi \& Bedding (2016), who introduced the undersampling correction and trialled Markov chain Monte Carlo (MCMC) software to solve binary orbits whilst providing uncertainties.

\begin{figure}[t]
\centering
\includegraphics[width=0.32\linewidth]{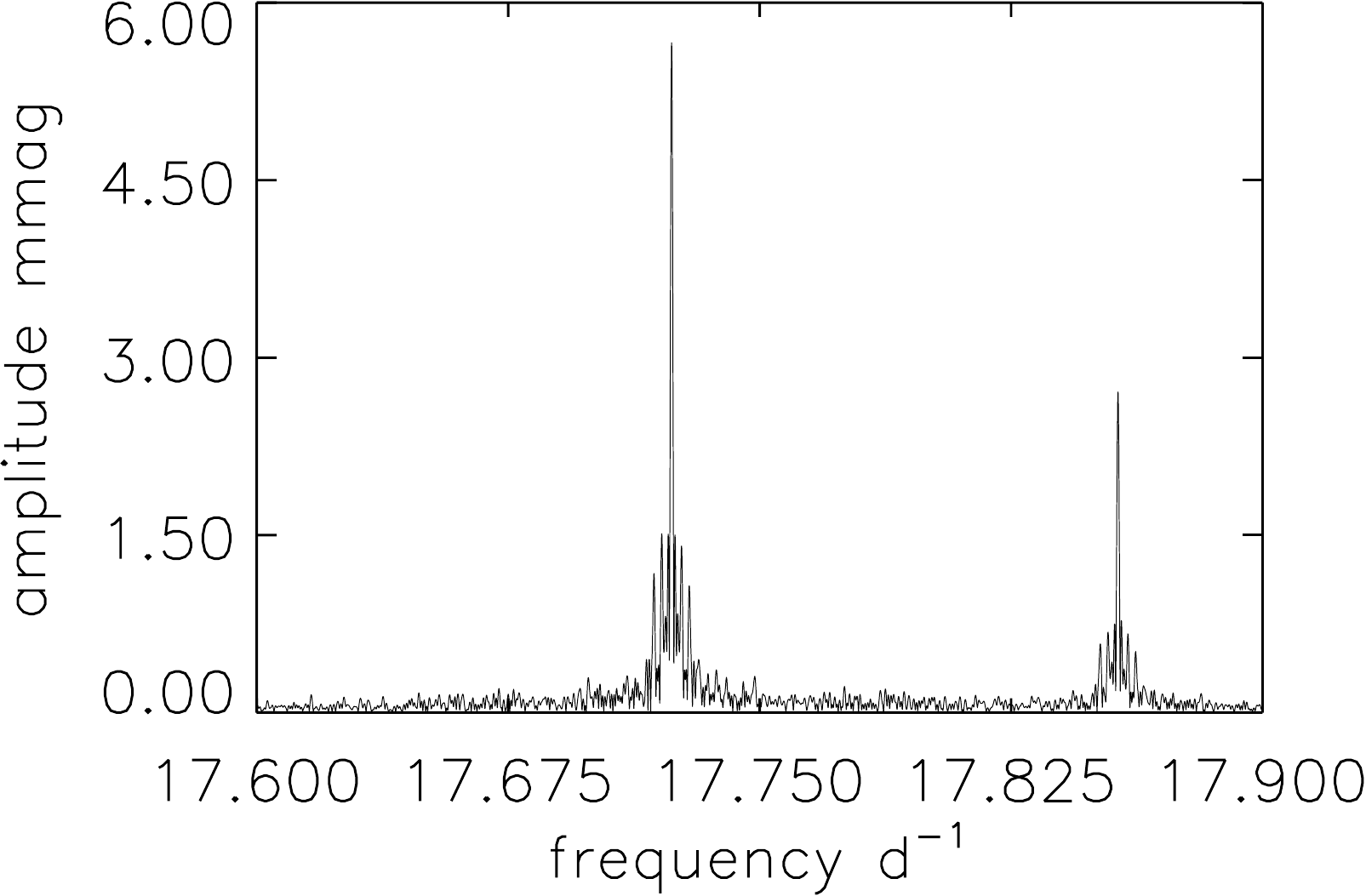}
\includegraphics[width=0.32\linewidth]{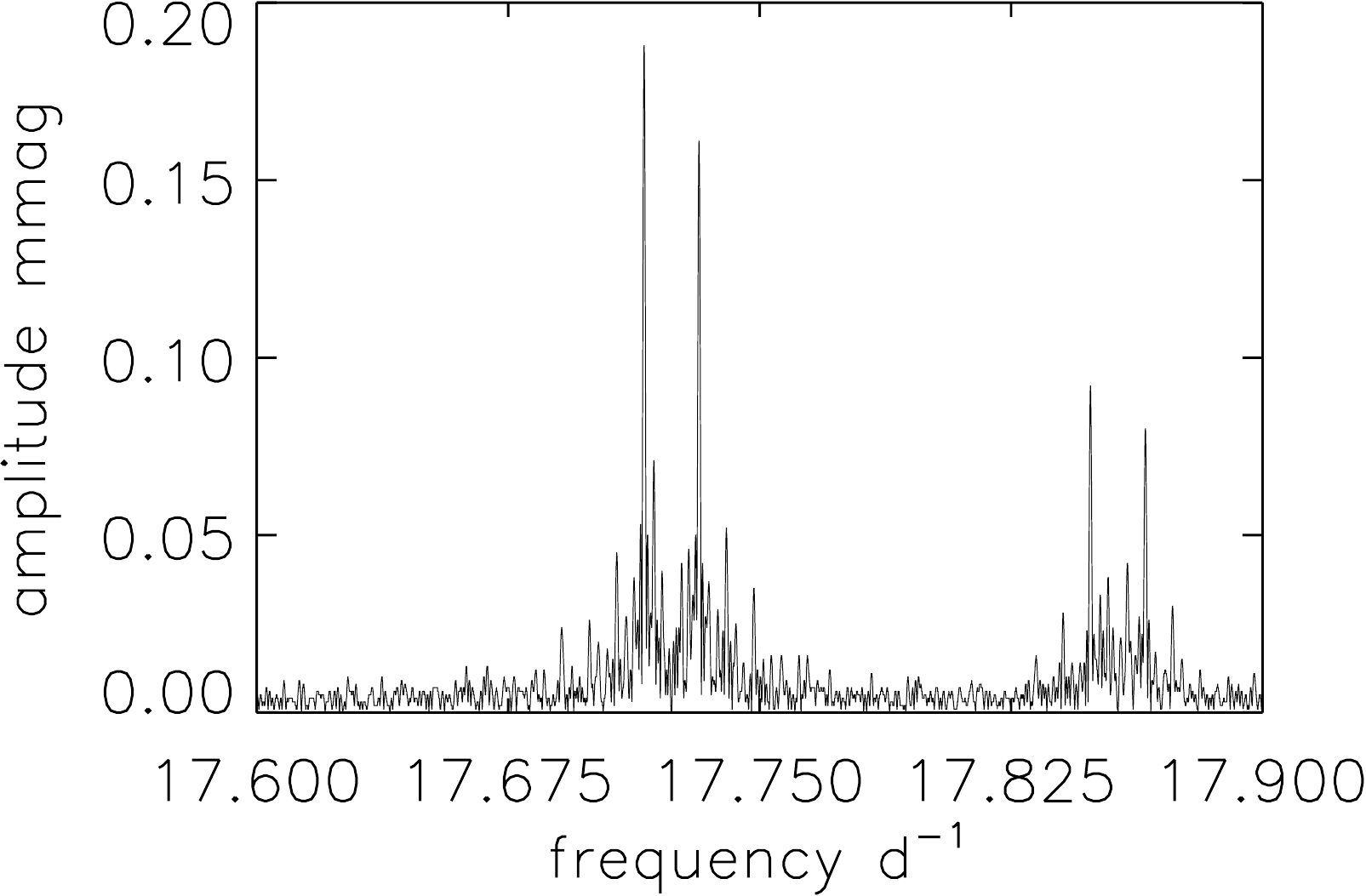}\\
\includegraphics[width=0.32\linewidth]{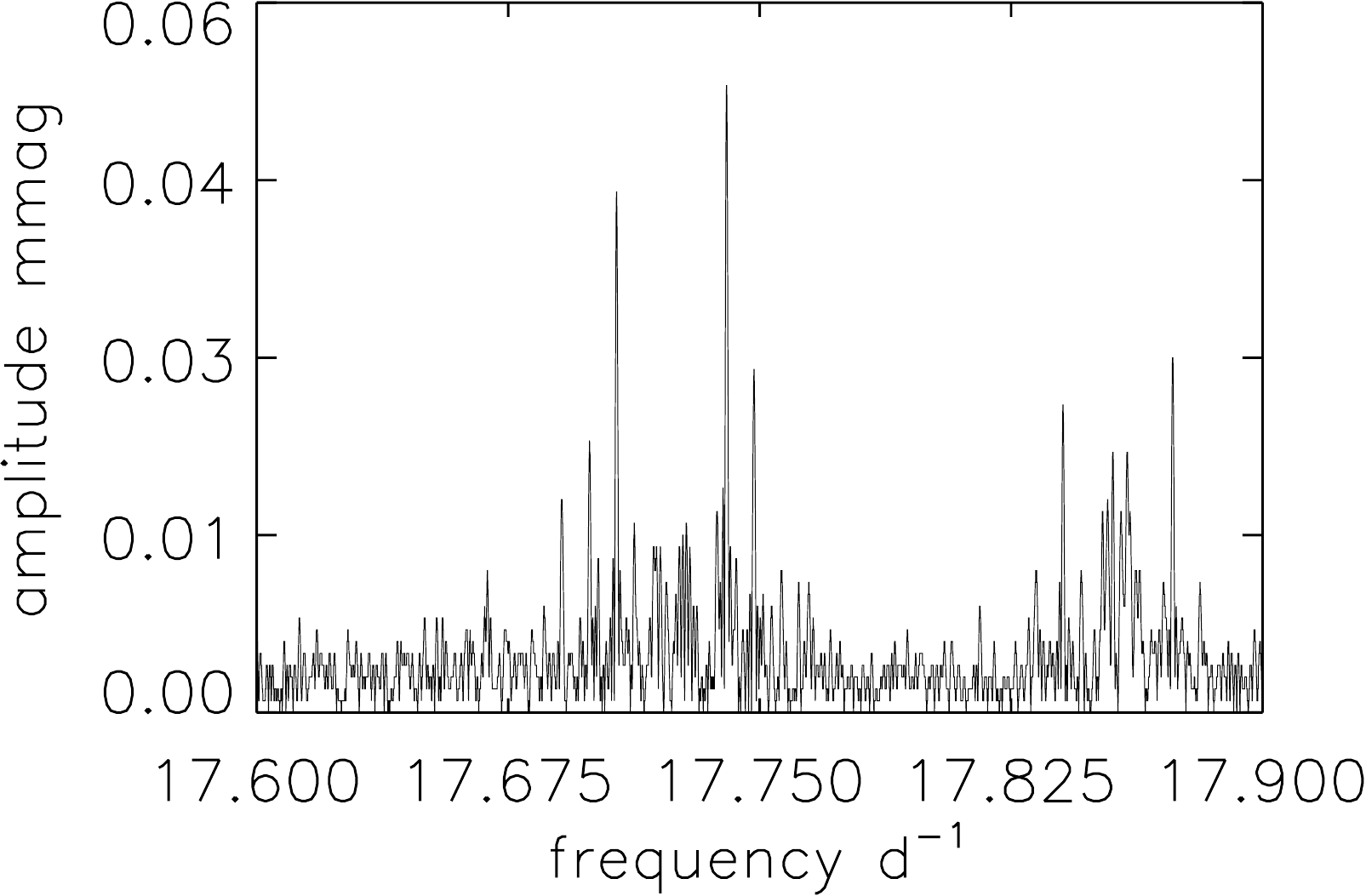}
\includegraphics[width=0.32\linewidth]{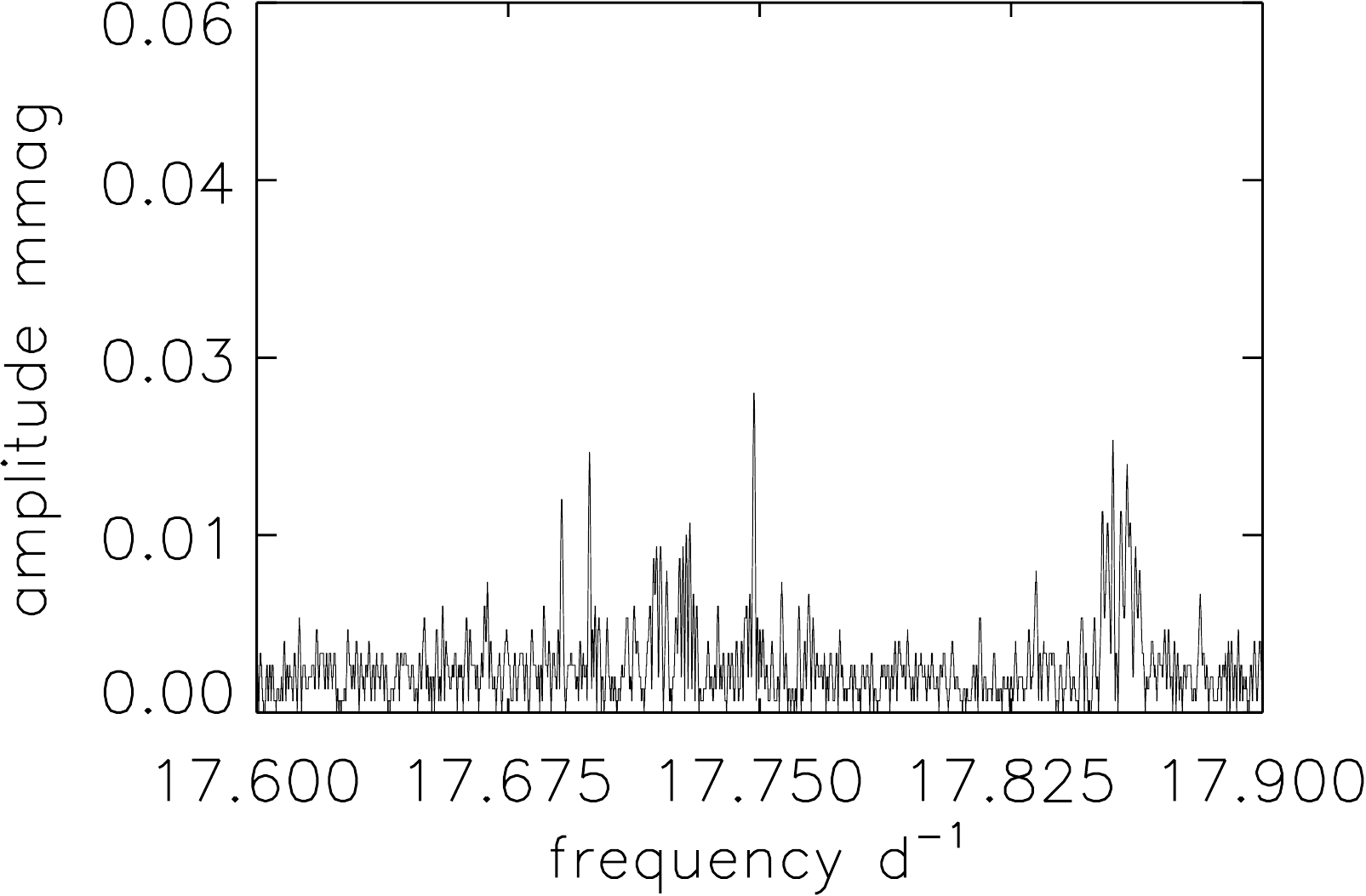}
\caption{An example of the Fourier transform of the light curve of a pulsating star (KIC\,10990452) in a binary system.
Adopted from Shibahashi et al. (2015).
\label{fig:02}}
\end{figure}

While the above method works very well to analyse binary stars with fairly long orbital periods, it is unfavourable to deal with binary stars with orbital periods shorter than the segment size dividing the observational time span, which is typically  $\sim$10\,d. A complementary method is to analyse the time sequence data in the Fourier domain instead of the time domain.
The phase variation of each mode caused by the light-time effect manifests itself, in the Fourier transform, as a multiplet with spacing equal to the orbital frequency as far as pulsations are coherent over the orbital period:
\begin{equation}
	\Delta L(t)
	=
	\Re
	\sum_k \left[
	\left\{
	{\cal A}^{(k)}_{0}{\rm e}^{2\uppi{\rm i}\nu_k t}
	+
	\sum_{m=1}^\infty
	\left(
	{\cal A}^{(k)}_{+m} {\rm e}^{2\uppi{\rm i}(\nu_k+m\nu_{\rm orb}) t}
	+
	{\cal A}^{(k)}_{-m} {\rm e}^{2\uppi{\rm i}(\nu_k-m\nu_{\rm orb}) t}	
	\right)
	\right\}
	\right],
\end{equation}
where ${\cal A}^{(k)}_0$ and ${\cal A}^{(k)}_{\pm m}$ are the complex amplitudes of the central peak of the mode $k$ and the sidelobes from the central component by $\pm m\nu_{\rm orb}$, respectively.
Figure\,\ref{fig:02} shows an example of the Fourier transform of the light curve obtained by {\it Kepler}, where the multiplet structure is unveiled iteratively.
The top left panel shows a zoom-in plot of the amplitude spectrum for KIC\,10990452 in the frequency range of the two highest amplitude p-mode peaks. After pre-whitening by the two highest peaks, the first  sidelobes become conspicuous, as seen in the top right panel. The bottom left panel shows that after further pre-whitening by the first sidelobes, the second and third sidelobes become easily visible.
In the bottom right panel the third sidelobes are easily seen for both main frequencies, and the low-frequency fourth sidelobe is also seen for the highest peak group.

The amplitudes and phases of the peaks in these multiplets reflect the orbital properties, hence the orbital parameters $(a_1\sin i, e, \varpi, \nu_{\rm orb}, t_{\rm p})$ can be extracted by analysing such multiplet structures in the Fourier domain. 
It should be verified that the frequency separation is the same and the multiplet structures are consistent with each other for different pulsation modes within the formal uncertainties.

Since in this approach the Fourier transform is carried out by using all of the data at once (that is, without dividing it into short segments), this method does not suffer from undersampling and the effects of poor frequency resolution, rendering it better suited to dealing with short-period binary systems. The shorter the orbital period, the larger the frequency separation of the multiplets and the easier the detection. 
On the other hand, the Fourier transform is unfavourable for binaries with orbital periods as long as the observational time span. In this sense, the time domain analysis and the Fourier domain analysis are complementary to each other. 
In the case that the both methods are applicable, the results obtained by the two methods have been shown to be in satisfactorily good agreement, as expected (see, e.g. Shibahashi, Kurtz \& Murphy 2015).
As for the theoretical relations between the multiplet properties and the orbital parameters, and also as for a method for determining these parameters, readers are recommended to refer Shibahashi \& Kurtz (2012) and Shibahashi, Kurtz \& Murphy (2015) for further details.

\section{Validating the photometric derivation of orbital parameters}
\begin{figure}[t]
\centering
\includegraphics[angle=0,width=0.7\linewidth]{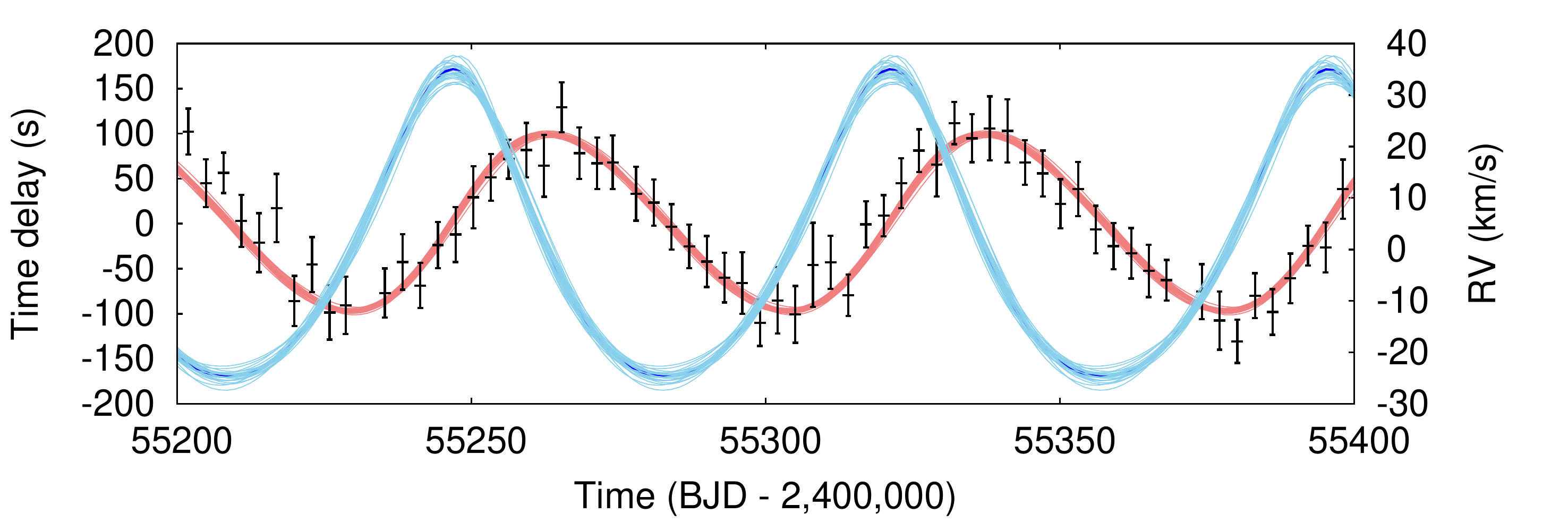}          
\includegraphics[angle=0,width=0.7\linewidth]{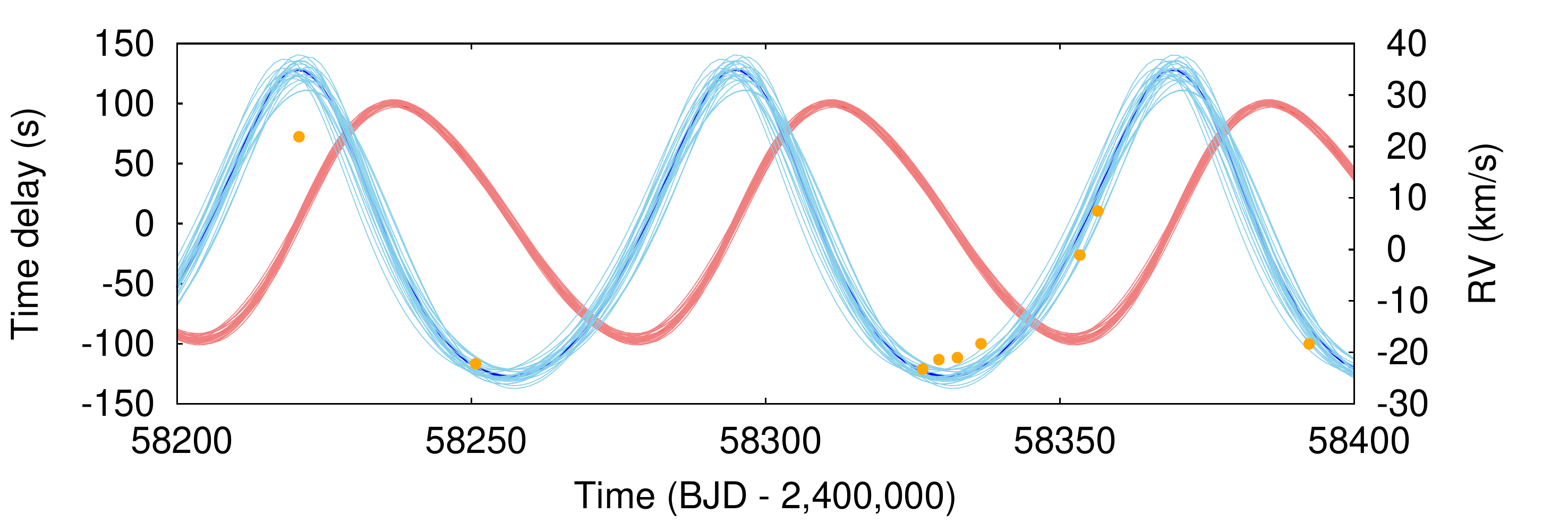}	
\caption{Top: The observed time delays with error bars derived from photometry for KIC\,1724961, with 25 random samples from the MCMC chain of the time-delay analysis shown as the red curve (left y-axes). The 25 corresponding RV curves are shown in blue (right y-axes). Only 200 of the 1470\,d of data are shown, but all 1470\,d were used (spanning the years 2009--2013).
Bottom: The RV curves are forward calculated for comparison with RV measurements taken several years later in 2018 (orange circles with error bars; the error bars are smaller than the plot symbols).}
\label{fig:03}
\end{figure}

Figure\,\ref{fig:03} demonstrates validation of the photometrically derived orbital parameters. In the top panel, the weighted average time delay data are shown with error bars, together with the best-fitting orbit computed using these time delays. 
The radial velocity is then predicted based on the orbital parameters.  To estimate the uncertainty, RV curves were generated with 25 sets of orbital parameters randomly selected from the Markov chain. The envelope of these curves is then regarded as the 96\% confidence level.
The actual spectroscopic determination of RV was carried out several years later. The measured RVs are shown with orange dots with error bars in the bottom panel. They are in good agreement with the photometric prediction. 
Note that the uncertainties on the measured RVs are smaller than those on the predicted curves in this case. As a consequence, combining the precise spectroscopic RVs and the photometric time delays taken several years apart substantially improves precision in the orbital parameters. A good example is seen in Derekas et al. (2019) for the 95-d binary KIC\,5709664. 
In particular, additional RV measurements significantly refine the orbital solutions when the orbital period is longer than the {\it Kepler} data set.

Validation of these new asteroseismic techniques for determination of the binary orbital elements
has also been demonstrated by comparing the results with those obtained by traditional eclipse timing for eclipsing binaries (see, e.g. Kurtz et al. 2015).

\section{Statistical study of binaries by asteroseismology}

\begin{figure}[t]
\centering
\includegraphics[angle=0,width=0.45\linewidth]{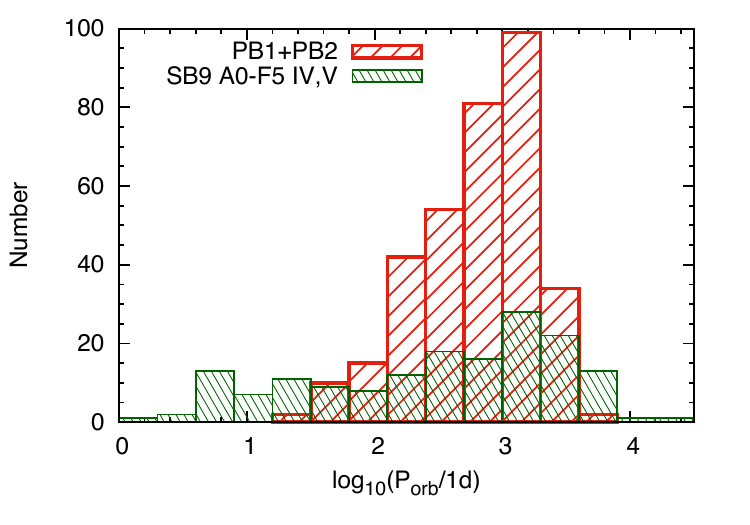}
\includegraphics[angle=0,width=0.45\linewidth]{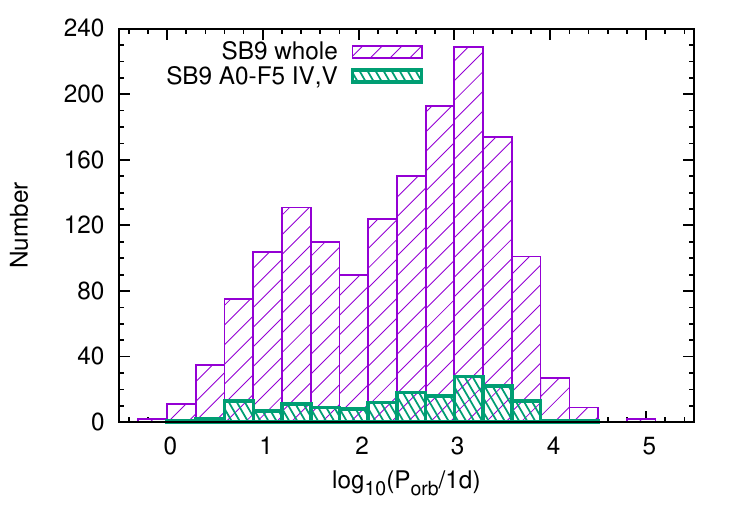}
\caption{Left: The distribution in orbital period of binaries that were asteroseismically found by Murphy et al. (2018). For comparison, spectroscopic binaries with primary stars in the similar temperature range (spectral type A0-F5 with luminosity class IV and V) listed in the  ninth catalogue of spectroscopic binary  (SB9) (Pourbaix et al. 2004)  are shown. 
Right: The distribution of all the binaries with full orbital solution listed in SB9 and the subset with primary stars with spectral type A0-F5 and luminosity class IV and V. }
\label{fig:04}
\end{figure}

Murphy et al. (2018) applied the pulsation phase method in time domain to light curves of all targets in the original {\it Kepler} view field with effective temperatures range 6600 and 10000\,K, aiming at capturing all $\delta$ Sct type pulsating stars in the instability strip. These targets exist in large numbers and also have coherent (phase-stable) oscillations. They adopted 10-day segmentation, hence binaries with orbital periods shorter than 20\,d were not found. With short-period binaries also having smaller orbits (hence smaller light travel times), the binaries with periods in the range of 20--100\,d are difficult to detect and the sample must be considerably incomplete. 
Although {\it Kepler} light curves were limited to 4\,yr in duration, some binaries with periods longer than this were still detected but the orbital solutions were multi-modal. RV data can help to resolve degeneracies in the solutions. In total, they investigated 12649 stars, of which 2224 main-sequence A and F stars were found to be suitable targets, based on their pulsation properties. Of those, 320 stars were found to be non-eclipsing single-pulsator binaries and 21 stars were found as double-pulsator binaries\footnote{Murphy et al. (2018) listed the orbital elements for 24 double-pulsator binaries. We reclassify, however, three stars among them as single-pulsator binaries in the present paper, since their confidential levels turned out to be low.}. In addition, 439 eclipsing binaries / ellipsoidal variables were also identified for future work.
The distribution of these 320+21 stars in orbital period is shown in the left panel of Fig.\,\ref{fig:04}, together with, for comparison,  
162 spectroscopic binaries listed in the ninth catalogue\footnote{http://sb9.astro.ulb.ac.be} of spectroscopic binaries (Pourbaix et al. 2004) with primary stars of similar spectral type (A0-F5) and luminosity class (IV to V). From the latter, only systems with full orbital solutions with uncertainties were selected. Stars without luminosity classes were removed.
The asteroseismically detected binaries tripled the number of intermediate-mass binaries with full orbital solutions, and importantly, provided a homogeneous data set for statistical analysis.

It has been noticed that the binary fraction of main-sequence primaries across the intermediate orbital period range is substantially larger in the case of OB primaries than the solar-type primaries (Abt, Gomez \& Levy 1990, Sana et al. 2013, Moe \& Di Stefano 2017). The former is around 35\,$\pm$\,12\% for O primaries and 23\,$\pm$\,7\% for mid-B primaries, but for solar-type primaries it is only about 5\%. The transition between these two regimes was, however, not well constrained. The binary fraction of the $\delta$\,Sct primaries (A-F main-sequence stars) in Murphy et al. (2018) was found to be (320\,+\,21)\,/\,2224\,=\,15.3\%, filling the gap in the statistics. This allowed Murphy et al. (2018) to parametrize a trend of increasing binary fraction proportional to $\log m_1$ for $m_1 \gtrsim 1\,{\rm M}_\odot$ while the fraction is nearly constant for $m_1 \lesssim 1\,{\rm M}_\odot$. 

\begin{figure}[t]
\centering
\includegraphics[width=0.45\linewidth]{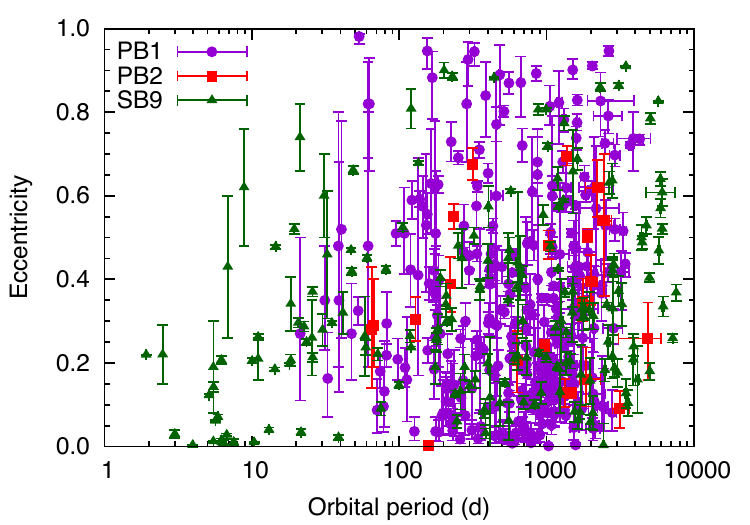}
\includegraphics[width=0.45\linewidth]{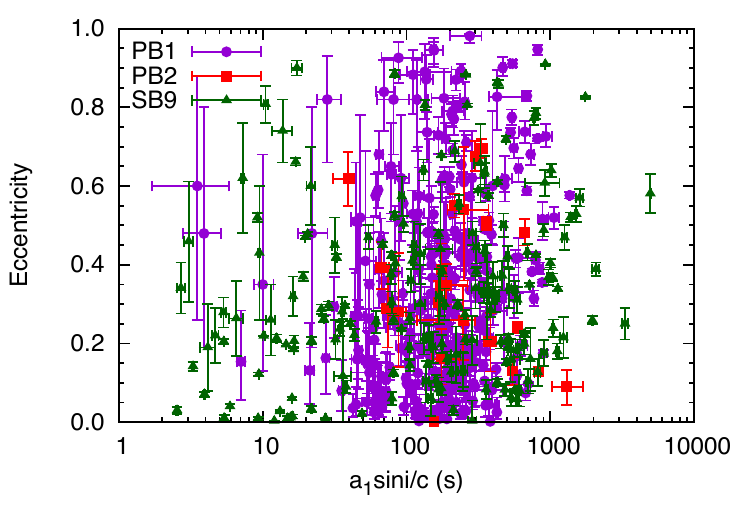}
\caption{Left: The orbital periods and eccentricities of the single-pulsator binaries (violet circles), double-pulsator binaries (red squares), and single-line spectroscopic binaries (open triangles). Right: The projected orbital semi-major axes  and eccentricities based on Murphy et al. (2018).
\label{fig:05}
}
\end{figure}

The period-eccentricity and the orbital size-eccentricity diagrams are shown in Fig.\,\ref{fig:05} with the addition of the 162 spectroscopic binaries. The uncertainties for all asteroseismically deduced parameters are from the time-delay analyses by Murphy et al. (2018). 
As they pointed out, at periods shorter than 150\,d, Fig.\,\ref{fig:05} shows that few systems have high eccentricities, $e\gtrsim0.7$, while there are many systems with low eccentricities, $e\lesssim0.3$. This fact implies that circularization at short periods is more efficient than expected.

\begin{figure}[t]
\centering
\includegraphics[width=0.55\linewidth]{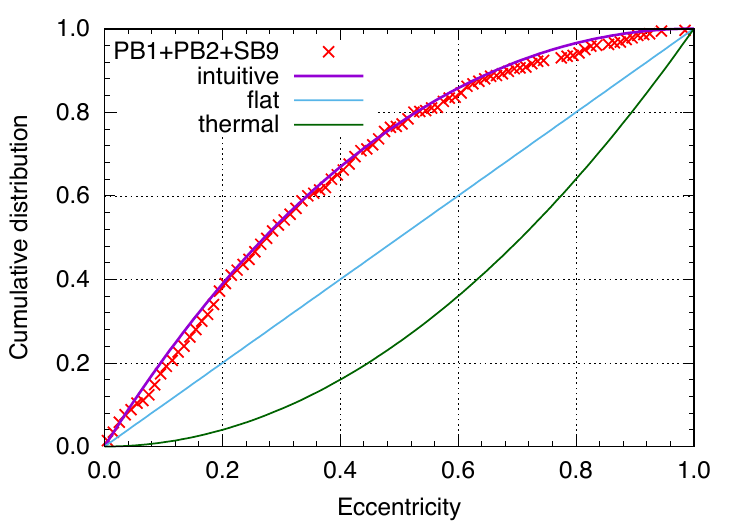}
\caption{Cumulative distribution of eccentricities of the current samples. It differs significantly from the thermal distribution, while it fits well the `intuitive' form adopted from Shen \& Turner (2008) with the parameter $a\rightarrow 1.0$.
\label{fig:06}
}
\end{figure}

Figure\,\ref{fig:06} shows the cumulative distribution of eccentricities of the current samples. 
It was suggested first by Jeans (1919) that the orbits of binary systems would reach a final state of equipartition of energy as a consequence of gravitational interaction with other stars. Distribution of eccentricities in the range of $[e, e+{\rm d}e]$ of such an equipartition state should be $2e\,{\rm d}e$ (Jeans 1919, see also Ambartsumian 1937\footnote{The paper was originally written in Russian and it is hard to find an original copy. Its English translation by D. W. Goldsmith (http://www.maths.ed.ac.uk/\url{~}douglas/Ambartsumian1937001.pdf) may be helpful for most readers.}, Kroupa 2008). 
As seen in Fig.\,\ref{fig:06}, the observed distribution in the current samples significantly differs from the equipartition state.  
It has been also known that the actual distribution of eccentricities of solar-type stars significantly deviates from the equipartition state. These facts imply that main-sequence binaries later than A-type do not form entirely through dynamical processing, which would lead to the equipartition state. 
Murphy et al. (2018) suggested that  the natal discs significantly moderate the eccentricities early in their evolution.

\begin{figure}[t]
\centering
\includegraphics[width=0.45\linewidth]{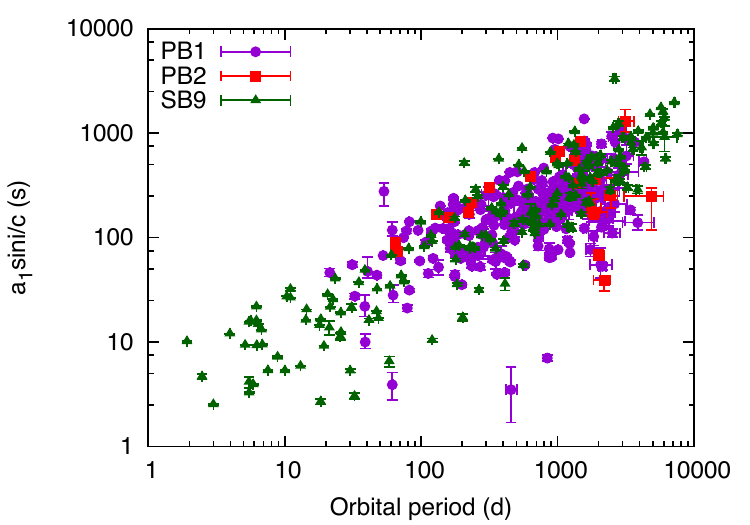}
\includegraphics[width=0.45\linewidth]{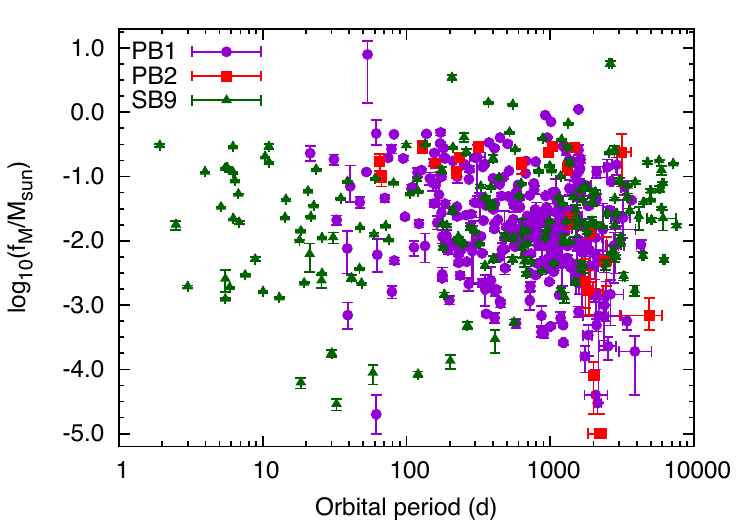}
\includegraphics[width=0.45\linewidth]{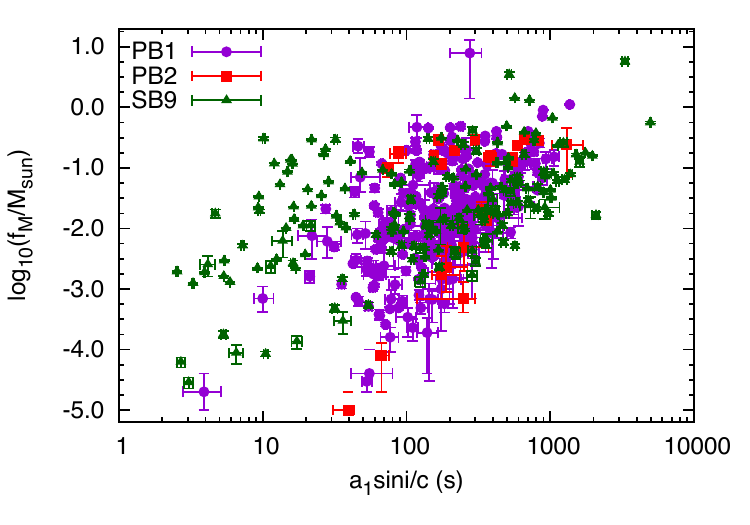}
\includegraphics[width=0.45\linewidth]{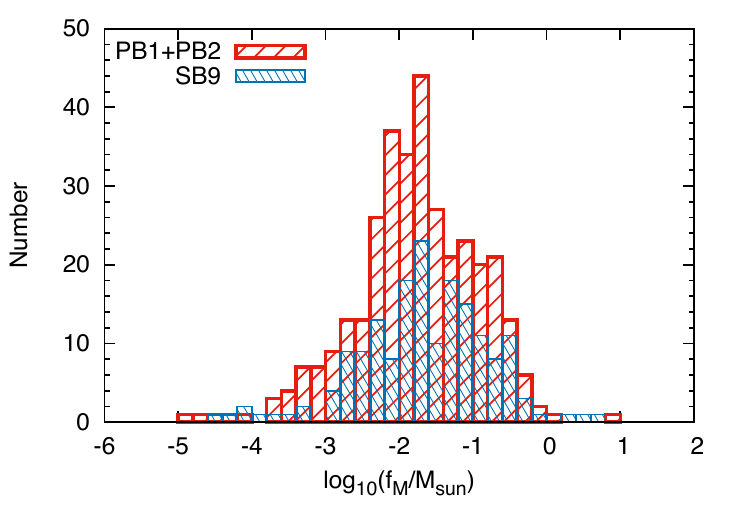}
\caption{Top left: The projected orbital sizes vs the orbital periods of the present samples, with the addition of the binaries listed in the ninth catalogue of spectroscopic binaries with similar properties. Top right: The mass functions vs the orbital periods. Bottom left: The mass functions vs the projected orbital sizes. Bottom right: Number distribution in the mass function.
\label{fig:07}
}
\end{figure}

Once the projected semi-major axis, $a_1\sin i$, and the orbital frequency, $\nu_{\rm orb}$, are deduced, the mass function, which gives the minimum mass of the companion, is determined:
\begin{eqnarray}
	f(m_1,m_2,\sin i) 
	&:=& 
	m_1{{q^3}\over{(1+q)^2}}\sin^3 i
	\\
	&=&{{4\pi^2 c^3}\over{G}} \nu_{\rm orb}^2 \left( {{a_1\sin i}\over{c}} \right)^3,
\end{eqnarray}
where $G$ denotes the gravitational constant and $q:=m_2/m_1$ is the mass ratio of the two components.
The top left panel of Fig.\,\ref{fig:07} shows the projected orbital sizes vs the orbital periods of the present samples. 
If we draw a line with the gradient equal to $2/3$ in this diagram, its $y$-intersect  specifies a mass function.
The top right and the bottom left panels of Fig.\,\ref{fig:07} show the mass function vs the orbital periods and the mass functions vs the projected orbital sizes of the present samples. 
The number distribution in $\log_{10}[f(m_1,m_2,\sin i)/{\rm M}_\odot]$ is shown in the bottom right panel.

The smallest mass function among the current samples, $5.3^{+1.1}_{-1.0}\times 10^{-7}\,{\rm M}_\odot$ (it is outside Fig.\,\ref{fig:07}), was found for KIC\,7917485, and a further careful analysis led to Murphy, Bedding \& Shibahashi (2016) concluding that the companion is a $12\,{\rm M}_{\rm Jup}$ planet orbiting in or near the habitable zone of the star with an orbital period of 840\,d.  
This is the first (and the only one yet) detection of an exoplanet via phase modulation of pulsations of the host star.

\section{An inverse problem to derive the mass-ratio distribution}
Analyses of pulsation phase variation tripled the number of intermediate-mass binaries with known mass functions. A more physically interesting quantity is, however, the mass-ratio distribution rather than the mass-function distribution. Murphy et al. (2018) parametrized the mass-ratio distribution using both inversion and Markov-chain Monte Carlo forward-modelling techniques after careful discussions about incompleteness of the observational sample, and found it to be skewed towards low-mass companions, peaking at $q \simeq 0.2$.  
Their result might have seemed quite different from the results of earlier investigations (e.g., Trimble 1987), most of which were based on statistics of visual binaries and double-line spectroscopic binaries. Those binary systems are advantageous for direct measurement of the mass ratio, but those with fainter companions are difficult to find so that the samples must be biased. Single-line spectroscopic binaries and pulsating stars showing phase variation due to orbital motion are sensitive to smaller mass-ratio binaries.

In this section, we try a different approach. We regard the inverse problem of getting the mass-ratio distribution from the known mass-function distribution as an inverse problem of an integral equation, and try to derive an analytical form of the solution. 
The main issue to overcome is the fact that the mass-function has the factor of $\sin i$, which cannot be observationally determined for individual systems. 

Let $\varphi$ and $\Phi$ be the cubic root of the mass function, $f$,
\begin{equation}
	\varphi 
	:= 
	f^{1/3},
\end{equation}
and the quantity defined as
\begin{equation}
	\Phi := m_1^{1/3} {{q}\over{(1+q)^{2/3}}} ,
\end{equation}
respectively, which are related to each other through
\begin{equation}
	\varphi = \Phi \sin i .
\end{equation}
We assume that the mass-function distribution is already known. Consequently, the distribution function of the quantity$\varphi$, denoted as $\Psi(\varphi)$, is a known function. Then, what is the distribution in $\Phi$?

\begin{figure}[t] 
\centering
\includegraphics[width=0.55\linewidth]{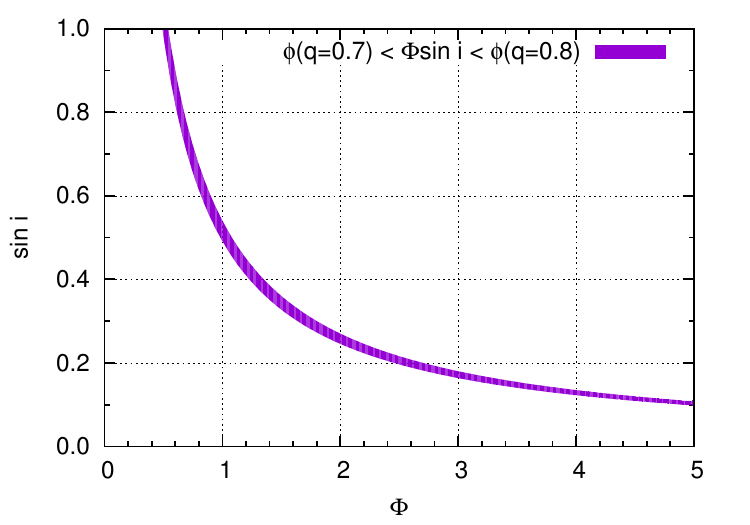}
\caption{An example of the integral range in the $(\Phi,\sin i)$-plane for equation (\ref{eq:09}).}
\label{fig:08}
\end{figure}

This problem is similar to that of determination of statistical distribution of stellar rotation velocities  from the observed projected velocities (Chandrasekhar \& M\"unch 1950).  
If the number of samples is large, it is reasonable to assume the orientations of binary orbital axes are distributed at random in space, i.e., the random distribution of the inclination angle, $i$.  
With this assumption,  
the probability that the inclination $i$ lies in the interval $[i, i+{\rm d}i]$ is given by $\sin i\,{\rm d}i$.
Let $p(\Phi)\,{\rm d}\Phi$ be the probability of occurrence of $\Phi$ in the range $[\Phi, \Phi+{\rm d}\Phi]$.
Since $\Phi$ and $i$ are independent from each other, the probability that $i$ and $\Phi$ appear in the range $[i, i+{\rm d}i]$ and  $[\Phi, \Phi+{\rm d}\Phi]$, respectively, is given by 
$p(\Phi)\sin i\,{\rm d}\Phi \,{\rm d}i$.
The probability that $\varphi$ falls in the fixed range $[\varphi, \varphi+{\rm d}\varphi]$ is then given by 
\begin{equation}
	\Psi(\varphi)\,{\rm d}\varphi 
	= 
	\int\!\!\!\int 
	p(\Phi)\sin i 
	\,{\rm d}\Phi\,{\rm d}i ,
\label{eq:09}
\end{equation}
where the right-hand side is a double integral with respect to $\Phi$ and the inclination angle $i$
with a constraint $\varphi \leq \Phi\sin i \leq \varphi+{\rm d}\varphi$. An example for the integral range in the $(\Phi,\sin i)$-plane is demonstrated in Fig.\,\ref{fig:08}.
Any point in this zone of the $(\Phi, \sin i)$-space lies on the same range of $[\varphi, \varphi+{\rm d}\varphi]$. 
This is a mapping of the one-dimensional space $[\varphi, \varphi+{\rm d}\varphi]$ on to the two-dimensional $(\Phi, \sin i)$-space\footnote{Note that the mapping with the constraint of $\Phi\sin i =\varphi$ is different from that with a fixed value of $i$. 
In the latter case, any value of $\varphi$ has a unique one-to-one correspondence of $\Phi$, while, in the case of the isotropic distribution of $i$, the value of $\Phi$ distributes along the curve shown in Fig.\,\ref{fig:08}.}. 

For a fixed value of $\Phi$, the integral range in $i$ is 
\begin{equation}
	{\rm d}i = {{{\rm d}\varphi}\over{\Phi\cos i}}.
\end{equation}
Therefore, equation (\ref{eq:09}) is rewritten as
\begin{equation}
	\Psi(\varphi)\,{\rm d}\varphi 
	= 
	{\rm d}\varphi \,\int_{\sin i=\varphi/\Phi} {{p(\Phi)\sin i}\over{\Phi\cos i}}\,{\rm d}\Phi,
\end{equation}
where the integral in the right-hand side should be performed with the constraint of $\sin i = \varphi/\Phi$.
By eliminating $\sin i$ and $\cos i$ in the integral 
with $\varphi$ and $\Phi$, we eventually get
\begin{equation}
	\Psi(\varphi) 
	= 
	\varphi \int_{\varphi}^\infty {{p(\Phi)}\over{\Phi \left(\Phi^2-\varphi^2\right)^{1/2}}}
	\,{\rm d}\Phi,
\label{eq:12}
\end{equation}
which is identical with the integral equation in Kuiper (1935).
Equivalent equations arise in a variety of problems of distributions of physical quantities observed only in projection (Craig \& Brown 1986).

Since the distribution function $\Psi(\varphi)$ is known, this equation is regarded as an integral equation with an unknown function to solve $p(\Phi)$, the distribution function of $\Phi$.
The problem is then turned to determine the unknown function $p(\Phi)$ from the known function $\Psi(\varphi)$ by solving the above integral equation.
Equation (\ref{eq:12}) is a case of Abel's integral equation, and the solution is analytically given by 
\begin{equation}
	p(\Phi)
	=
	-{{2}\over{\uppi}} \Phi^2 {{{\rm d}}\over{{\rm d}\Phi}}
	\left[\,
	\Phi \int_\Phi^\infty {{\Psi(\varphi)}\over{\varphi^2 \left(\varphi^2-\Phi^2\right)^{1/2}}} 
	\,{\rm d}\varphi
	\,\right]
\label{eq:13}
\end{equation}
(see Appendix).

\begin{figure}[t]
\centering
\includegraphics[width=0.45\linewidth]{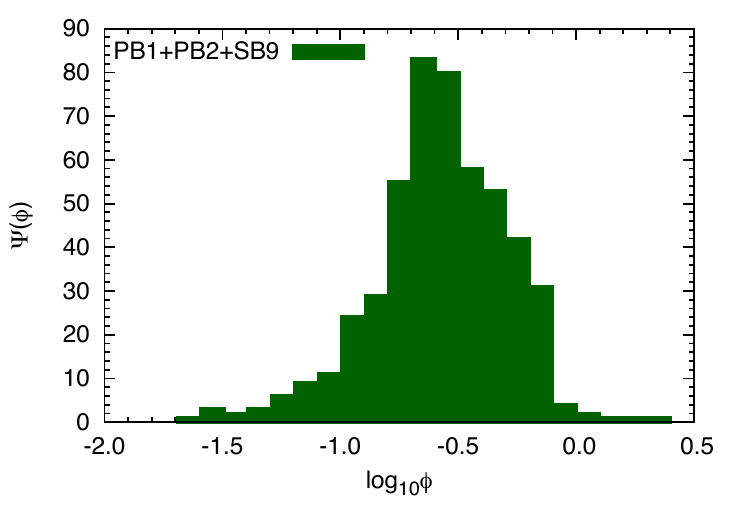}
\includegraphics[width=0.45\linewidth]{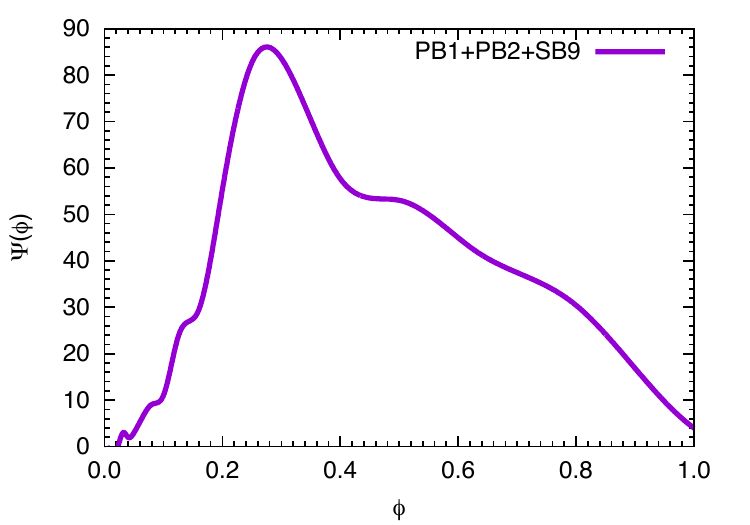}
\caption{Left: Histogram of the current samples in $\log_{10} \varphi$. 
Right: The smoothed distribution function $\Psi(\varphi)$, obtained with the cubic spline.
Note that the $x$-axis is  linear in $\varphi$ in the right panel, while decimal logarithmc in the left panel.}
\label{fig:09}
\end{figure}

The greater the sample size, the more reasonable our assumption of the random distribution in $i$ becomes. 
In estimating $\Psi(\varphi)$, we use all the stars discussed in the previous section together, 320+21 stars analysed from the {\it Kepler} data together with 162 stars in the ninth spectroscopic binary catalogue. 
The left panel of Fig.\,\ref{fig:09} shows the histogram of the current data with respect to $\log_{10}\varphi$, where the binning size is 0.1.
We convert it into a smooth function by using the cubic spline. The right panel of Fig.\,\ref{fig:09} shows the continuous function $\Psi(\varphi)$ thus obtained. 

\begin{figure}[htbp]
\centering
\includegraphics[width=0.45\linewidth]{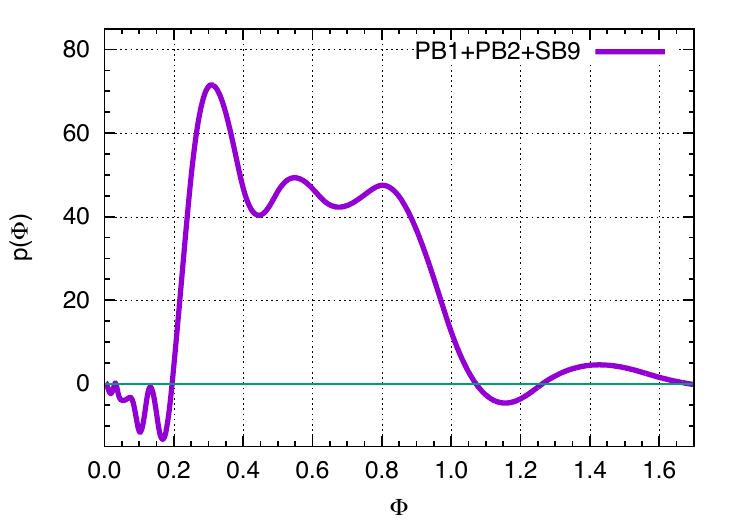}
\includegraphics[width=0.45\linewidth]{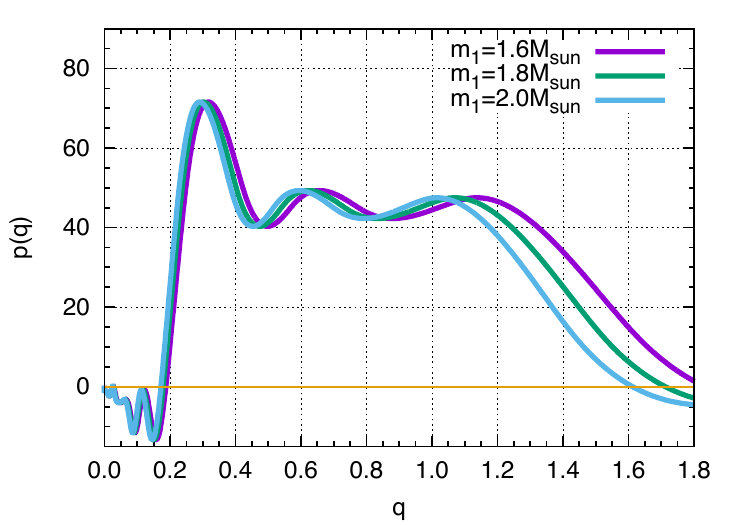}
\caption{Left panel: Solution $p(\Phi)$ of the integral equation with $\Psi(\varphi)$ shown in the right panel of Fig.\,\ref{fig:09}. 
Right panel: The number distribution $p(q)$ of the current samples as a function of the mass ratio $q$, converted from $p(\Phi)$, in the case of $m_1=1.6, 1.8$ and $2.0\,{\rm M}_\odot$.
}
\label{fig:10}
\end{figure}

Then, by computing numerically the right-hand side of equation (\ref{eq:13}), we get the distribution in $\Phi$, which is shown in the left panel of Fig.\,\ref{fig:10}.
As seen in this figure, the solution $p(\Phi)$ falls below zero in some range of $\Phi$, which is obviously unphysical. 
This is a direct consequence that the numbers of samples in the rages of $\varphi \lesssim 0.1$ and $\varphi \gtrsim 0.9$ are too small to justify our assumption about the random distribution in $i$.
This might be caused by the underlying distribution of the mass ratio itself (i.e.\ binaries with extreme mass ratios are rare), but certainly at small values of $\varphi$ the detection efficiency is small. 
We have not followed Murphy et al. (2018) in accounting for the selection biases. Instead we demonstrate here the inversion method using the present samples only, and we do not attempt to recover the underlying distribution of binary mass ratios.   
We consider the overall feature of $p(\Phi)$ peaking around $\Phi\simeq 0.3$ to be reliable,
but the inverted result $p(\Phi)$ in the range of $\Phi \lesssim 0.2$ and $\Phi \gtrsim 1.0$ should be regarded as unreliable.
   
We convert the distribution in $\Phi$ into that in the mass ratio, $q$, by supposing that the mass range of the primary components, $\delta$\,Sct-type pulsating stars and the A0-F5 stars mostly with the luminosity class IV and V, is $1.6\sim 2.0\,{\rm M}_\odot$.  
The final result is shown in the right panel of Fig.\,\ref{fig:10}.  

\section{Implications of the mass-ratio distribution}
Statistically small samples in the range of $\varphi \lesssim 0.1$ and $\varphi \gtrsim 0.9$ make the solution at $q \lesssim 0.2$ and $q \gtrsim 1.5$ unreliable. 
Therefore, we restrict ourselves to discussing the characteristics of the mass-ratio distribution in the range of $0.2 \lesssim q \lesssim 1.5$.
Conspicuous features of the solution $p(q)$ in this range are:
\begin{itemize}
\item the solution $p(q)$ 
has a peak around $q\simeq 0.3$, which corresponds to $m_2 \simeq 0.5\,{\rm M}_\odot$,
\item it is nearly flat in the range of $0.5 \lesssim q \lesssim 1.0$, which implies that the distribution in the companion mass, $m_2$, is nearly flat in the range of $0.8  \lesssim m_2/{\rm M}_\odot \lesssim 1.8$,
\item it gradually decreases with the increase in $q$ for $1.0 \lesssim q$.
\end{itemize}

As for the first point in the above, 
the original {\it Kepler} field lies out of the galactic plane, as pointed out in Murphy et al. (2018), 
and the most massive stars in the field must have already evolved off the main sequence and become compact objects.
Hence the peak around $q\simeq 0.3$ is likely to be attributed to the white dwarfs in addition to K-type main-sequence stars.
This may be observationally tested by using samples of more massive B stars, once available.

As for the second point, it is important to note that the both components of twenty-one double-pulsator binaries in our sample are $\delta$\,Sct type stars, hence their mass ratios are around $0.8 \lesssim q \lesssim 1.0$. This substantially contributes to enhance the fraction of this range. The spectroscopic binary systems also contribute heavily to this range of mass ratio, leading to a different shape than observed by Murphy et al. (2018).

Since A and early F stars are sampled in the present analysis, the secondary stars of the samples must be mostly F5- or late type main-sequence stars. If the secondary were a more massive main-sequence star, the system would likely have been classified with a hotter temperature and fallen outside the sample range. Hence the possibility of $q \gtrsim 1.0$ is mostly restricted to cases of compact unseen secondaries, which evolved from massive main-sequence stars. White dwarfs are excluded, as the Chandrasekhar limit is $\sim 1.4\,{\rm M}_\odot$, which is lower than the $\delta$\,Sct masses. The remaining possibility is then only neutron stars, which have a critical mass limit of $\sim 3\,{\rm M}_\odot$, or more massive black holes.
Since $p(q) \approx 0$ for $q \gtrsim 1.5$, the presence of a black hole in the current samples seems unlikely. On the other hand, the expected number of the systems with $1.3\lesssim q \lesssim 1.5$ is several among the current samples, so the probability is not excluded. 

\section{Future prospects}
Stellar evolution theory tells us that stars with an initial mass more than $\sim 30\,{\rm M}_\odot$ ultimately become black holes, so stellar-mass black holes should be ubiquitous. Nevertheless, only around 20 stellar-mass black holes have been dynamically confirmed to date (Cowley 1992, Casares 2007, Corral-Santana et al. 2016), and they have all been found through their X-ray emission and high energy physics, as either High Mass X-Ray Binaries (HMXB) whose optical counterparts are early type stars which are blowing gas to black holes, or soft X-ray transients among Low Mass X-Ray Binaries (LMXB) whose optical counterparts are late type stars filling their Roche lobes and transferring mass on to black holes. On the other hand, it is natural to expect stellar-mass quiet black holes without X-ray emission in binary systems with large separation, if these binaries survive the supernova kicks. The discovery of black holes in the optical through their gravitational interactions would be a major scientific breakthrough. 
As demonstrated in the present paper, space-based photometry has made it possible to measure the orbital phase variation or the frequency modulation of pulsating stars in binary systems with extremely high precision over long time spans. The lower limit for the mass of the binary companion is obtained by analysis of phase variation or frequency modulation of the pulsating star. If the companions are non-luminous and if the masses of the companions exceed the mass limit for neutron stars ($\sim 3\,{\rm M}_\odot$), the companions are likely to be black holes though follow-up RV observations from the ground are necessary to exclude some other possibilities, such as multiple systems composed of faint stars. 
This is the most efficient way of finding quiet stellar-mass black holes, making for attractive complementary science that can be done with precise space photometry with long time spans.

%
%
\section*{Acknowledgements}
It was our great pleasure to celebrate the 75th birthday of  a very respected colleague, and also our close friend, Arlette, together with all the participants and other many colleagues in the world. 
The authors thank the organizing committee of the workshop, led by Marc-Antoine Dupret, for their excellent organization and good atmosphere of this fruitful meeting.
This work was partly supported by the JSPS Grant-in-Aid for Scientific Research (16K05288).

%
%
%

\footnotesize
\beginrefer
\refer Abt H. A., Gomez A. E., Levy S. G. 1990, ApJS, 74, 551

\refer Ambartsumian V. 1937, Astron. Zh., 14, 207

\refer Auvergne M., Bodin P., Boisnard L. et al. 2009, A\&A, 506, 411

\refer Balona L. A. 2014, MNRAS, 443, 1946

\refer Barnes T. G. III, Moffett T. J. 1975, AJ, 80, 48

\refer Casares J. 2007, in Black Holes from Stars to Galaxies --- Across the Range of Masses, Proceedings of IAU Symposium 238, 3

\refer Chandrasekhar S., M\"unch G. 1950, ApJ, 111, 142

\refer Corral-Santana J. M., Casares J., Mu\~{n}oz-Darias T., Bauer F. E., Mart\'{\i}nez-Pais I. G., Russel D. M. 2016, A\&A, 587, A61

\refer Cowley A. P. 1992, ARA\&A, 30, 287

\refer Craig I. J. D., Brown J. C. 1986, Inverse Problems in Astronomy, Adam Hilger Ltd.

\refer Derekas A., Murphy S. J., D\'alya G. et al. 2019, MNRAS, 486, 2129

\refer Duch\^{e}ne G., Kraus A. 2013, ARA\&A, 51, 269

\refer Duquennoy A., Mayor M. 1991, A\&A, 248, 485

\refer Hulse R. A., Taylor J. H. 1975, ApJ, 195, L51

\refer Jeans J. H. 1919, MNRAS, 79, 408

\refer Kirk B., Conroy K., Pr\v{s}a A. et al. 2016, AJ, 151, 68

\refer Kobulnicky H. A., Kiminki D. C., Lundquist M. J. et al. 2014, ApJS, 213, 34

\refer Koch D. G., Borucki W. J., Basri G. et al. 2010, ApJ, 713, L79

\refer Koen C. 2014, MNRAS, 444, 1486

\refer Kroupa P. 2008, in The Cambridge N-Body Lectures, Lecture Notes in Physics, 760, 181

\refer Kuiper G. P. 1935, PASP, 47, 15

\refer Kurtz D. W., Hambleton K. M., Shibahashi H., Murphy S. J., Pr\v{s}a A. 2015, MNRAS, 446, 1223

\refer Moe M, Di Stefano R. 2017, ApJS, 230, 15

\refer Moffett T. J., Barnes T. G. III, Fekel F. C. Jr., Jefferys W. H., Achtermann J. M. 1988, AJ, 95, 1534

\refer Murphy S. J., Shibahashi H., Kurtz D. W. 2013, MNRAS, 430, 2986

\refer Murphy S. J., Bedding T. R., Shibahashi H., Kurtz D. W., Kjeldsen H. 2014, MNRAS, 441, 2515

\refer Murphy S. J., Shibahashi H. 2015, MNRAS, 450, 4475

\refer Murphy S. J., Shibahashi H., Bedding T. R. 2016, MNRAS, 461, 4215

\refer Murphy S. J., Bedding T. R., Shibahashi H. 2016, ApJ, 827, L17

\refer Murphy S. J., Moe M., Kurtz D. W., Bedding T. R., Shibahashi H., Boffin H. M. J. 2018, MNRAS, 474, 4322

\refer Murphy S. J. 2018, arXiv:1811.12659

\refer Pourbaix D., Tokovinin A. A., Batten A. H. et al. 2004, A\&A, 424, 727

\refer Raghavan D., McAlister H. A., Henry T. J. et al. 2010, ApJS, 190, 1

\refer Ricker G. R., Winn J. N., Vanderspek R. et al. 2015, Journal of Astronomical Telescopes, Instruments, and Systems, 1, 014003

\refer Sana H., de Mink S. E., de Koter  A. et al. 2012, Science, 337, 444

\refer Shen Y., Turner E. L. 2008, ApJ, 685, 553

\refer Shibahashi H., Kurtz D. W. 2012, MNRAS, 422, 738

\refer Shibahashi H., Kurtz D. W., Murphy S. J. 2015, MNRAS, 450, 3999

\refer Shibahashi H., Murphy S. J. 2018, arXiv:1811.10205

\refer Silvotti R., Schuh S., Janulis R. et al. 2007, Nature, 449, 189

\refer Sterken C. 2005, The Light-Time Effect in Astrophysics: Causes and Cures of the O\,--\,C Diagram, Proceedings of ASP Conference Series, 335

\refer Trimble V. 1987, AN, 308, 343

\refer Walker G., Matthews J., Kusching R. et al. 2003, PASP, 115, 1023

\refer Weiss W. W., Rucinski S. M., Moffat A. F. J. et al. 2014, PASP, 126, 573

\refer Wolszczan A., Frail D. A. 1992, Nature, 355, 145

\endrefer

\normalsize
\renewcommand{\theequation}{A\arabic{equation}}
\setcounter{equation}{0}
\renewcommand{\thefigure}{A\arabic{figure}}
\setcounter{figure}{0}
\section*{Appendix: Derivation of equation (\ref{eq:13})}
Converting the variables $\Phi$ and $\varphi$ to the new variables $\zeta$ and $\eta$ defined by
\begin{equation}
	\Phi =: \zeta^{-1/2}
\label{eq:A1}
\end{equation}
and
\begin{equation}
	\varphi =: \eta^{-1/2},
\label{eq:A2}
\end{equation}
respectively,
and rewriting formally
\begin{equation}
	g(\eta) := \Psi(\varphi)
\label{eq:A3}
\end{equation}
and
\begin{equation}
	f(\zeta) := {{\Phi}\over{2}}  \, p(\Phi),
\label{eq:A4}
\end{equation}
we rewrite equation (\ref{eq:12}) with the following form:
\begin{eqnarray}
	g(\eta)
	&=&
	\eta^{-1/2}
	\int_\eta^0 
	{{2f(\zeta)}\over{(\zeta^{-1}-\eta^{-1})^{1/2}}}
	{{1}\over{\Phi^2}}{{{\rm d}\Phi}\over{{\rm d}\zeta}}\,{\rm d}\zeta
	\nonumber\\
	&=&
	\int_\eta^0 
	{{f(\zeta)}\over{(\eta - \zeta)^{1/2}}}
	\,{\rm d}\zeta.	
\label{eq:A5}
\end{eqnarray}

\begin{figure}[t]
\begin{center}
	\includegraphics[width=0.7\linewidth]{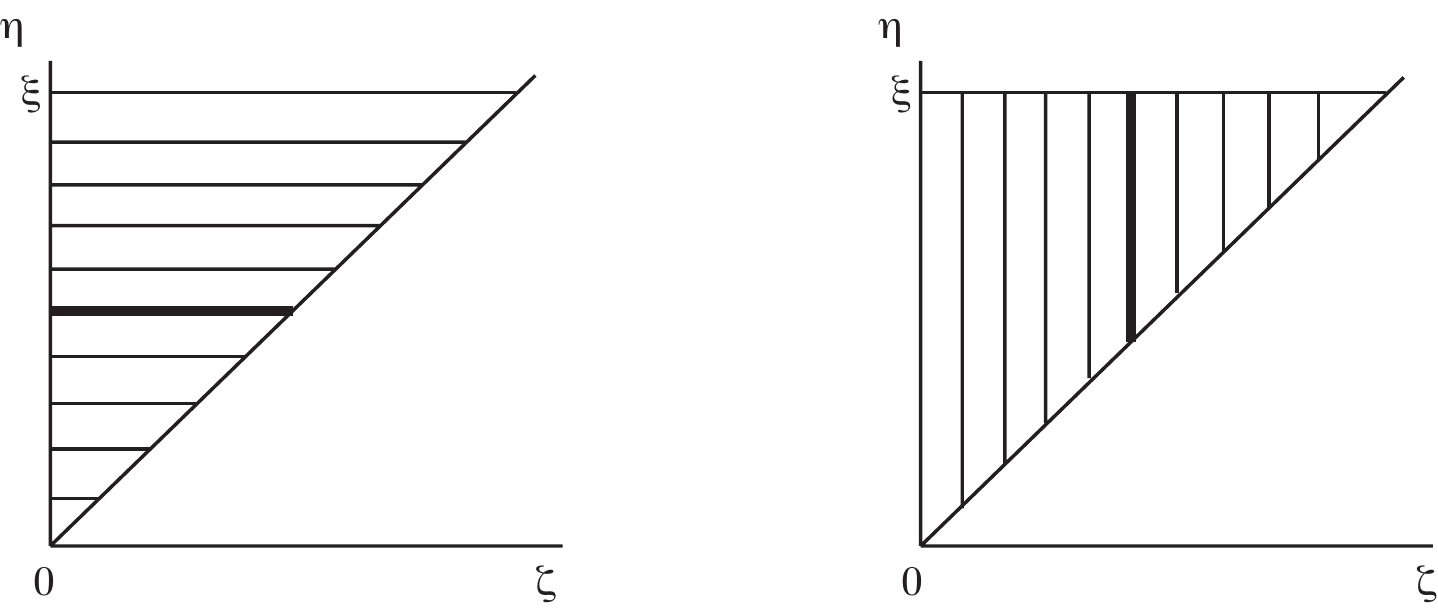}
	\caption{Two ways of the two-dimensional integral on the $(\zeta, \eta)$-plane. The left panel illustrates the way of the integral shown in the left-hand side of equation (\ref{eq:A6}), while the right panel illustrates the integral shown in the right-hand side of equation (\ref{eq:A7}). }
	\label{fig:A1}
\end{center}
\end{figure}

Multiplying $(\xi-\eta)^{-1/2}$ with the both sides of equation (\ref{eq:A5}) and then integrate from $0$ to $\xi$ with respect to $\eta$, we get
\begin{equation}
	\int_0^\xi {{g(\eta)}\over{(\xi-\eta)^{1/2}}} \,{\rm d}\eta
	=
	\int_0^\xi {{1}\over{(\xi-\eta)^{1/2}}} 
	\left[
	\int_0^\eta  {{f(\zeta)}\over{(\eta-\zeta)^{1/2}}} \,{\rm d}\zeta
	\right]\, {\rm d}\eta .
\label{eq:A6}
\end{equation}
Changing the order of integrals (see Fig.\,\ref{fig:A1}), the right-hand side of equation (\ref{eq:A6}) is reduced to
\begin{equation}
	\int_0^\xi {{1}\over{(\xi-\eta)^{1/2}}} 
	\left[
	\int_0^\eta  {{f(\zeta)}\over{(\eta-\zeta)^{1/2}}} \,{\rm d}\zeta
	\right]\, {\rm d}\eta 
	=
	\int_0^\xi f(\zeta) 
	\left[
	\int_\zeta^\xi 
	{{{\rm d}\eta}\over{(\xi-\eta)^{1/2} (\eta-\zeta)^{1/2}}} 
	\right]\,{\rm d}\zeta.
\label{eq:A7}
\end{equation}
Note that
\begin{equation}
	\int_\zeta^\xi 
	{{{\rm d}\eta}\over{(\xi-\eta)^{1/2} (\eta-\zeta)^{1/2}}} 
	= \uppi.
\label{eq:A8}
\end{equation}
Then
\begin{equation}
	\int_0^\xi {{g(\eta)}\over{(\xi-\eta)^{1/2}}} \,{\rm d}\eta
	=
	\uppi \int_0^\xi f(\zeta) \,{\rm d}\zeta,
\label{eq:A9}
\end{equation}
hence
\begin{equation}
	f(\zeta) = 
	{{1}\over{\uppi}} 
	{{{\rm d}}\over{{\rm d}\zeta}}
	\left[ \int_0^\zeta {{g(\eta)}\over{(\zeta-\eta)^{1/2}}} \,{\rm d}\eta	\right].
\label{eq:A10}
\end{equation}
Changing back to the original variables $(\varphi, \Phi)$ from $(\zeta, \eta)$,
we obtain
\begin{equation}
	p(\Phi) = -{{2}\over{\uppi}} \Phi^2 
	{{{\rm d}}\over{{\rm d}\Phi}} 
	\left[
	\Phi 
	\int_\Phi^\infty {{\Psi(\varphi)}\over{\varphi^2 \left(\varphi^2-\Phi^2\right)^{1/2}}} \,{\rm d}\varphi
	\right].
\label{eq:A11}
\end{equation}

\end{document}